\documentclass[a4paper,12pt]{article}
\pdfoutput=1
\usepackage{graphicx, rotating,amssymb,amsmath,soul}

\ifx\pdfoutput\undefined
\usepackage[dvips,bookmarks]{hyperref}	
\else
\usepackage{hyperref}	
\fi
\hypersetup{colorlinks,bookmarksopen,bookmarksnumbered,citecolor=rossoc,
linkcolor=blue,pdfstartview=FitH,urlcolor=rossos}
\def\myurl#1#2{\href{http://#1}{#2}}
\def\hhref#1{\href{http://arxiv.org/abs/#1}{#1}} 

\usepackage{multicol}
\usepackage{color}
\definecolor{rosso}{cmyk}{0,1,1,0.4}
\definecolor{rossos}{cmyk}{0,1,1,0.55}
\definecolor{rossoc}{cmyk}{0,1,1,0.2}
\definecolor{blu}{cmyk}{1,1,0,0.3}
\definecolor{blus}{cmyk}{1,1,0,0.6}
\definecolor{bluc}{cmyk}{1,1,0,0.1}
\definecolor{verde}{cmyk}{0.92,0,0.59,0.25}
\definecolor{verdec}{cmyk}{0.92,0,0.59,0.15}
\definecolor{verdes}{cmyk}{0.92,0,0.59,0.4}

\font\tenrsfs=rsfs10 at 12pt
\font\sevenrsfs=rsfs7
\font\fiversfs=rsfs5
\newfam\rsfsfam
\textfont\rsfsfam=\tenrsfs
\scriptfont\rsfsfam=\sevenrsfs
\scriptscriptfont\rsfsfam=\fiversfs
\def\mathscr#1{{\fam\rsfsfam\relax#1}}

\def\circa#1{\,\raise.3ex\hbox{$#1$\kern-.75em\lower1ex\hbox{$\sim$}}\,}

\newcommand{\beq}{\begin{equation}}
\newcommand{\eeq}{\end{equation}}

\def\circa#1{\,\raise.3ex\hbox{$#1$\kern-.75em\lower1ex\hbox{$\sim$}}\,}
\makeatletter

\def\art{\@ifnextchar[{\eart}{\oart}}
\def\eart[#1]#2#3#4#5#6{{\rm #2}, {#3 #4} {\rm (#6) #5} [{\hhref{#1}}]}
\def\hepart[#1]#2{{\rm #2, \hhref{#1}}}
\newcommand{\oart}[5]{{\rm #1}, {#2 #3} {\rm (#5) #4}}

\newcounter{alphaequation}[equation]
\def\thealphaequation{\theequation\hbox to
0.6em{\hfil\alph{alphaequation}\hfil}}
\def\eqnsystem#1{
\def\@eqnnum{{\rm (\thealphaequation)}}
\def\@@eqncr{\let\@tempa\relax \ifcase\@eqcnt \def\@tempa{& & &} \or
  \def\@tempa{& &}\or \def\@tempa{&}\fi\@tempa
  \if@eqnsw\@eqnnum\refstepcounter{alphaequation}\fi
\global\@eqnswtrue\global\@eqcnt=0\cr}
\refstepcounter{equation} \let\@currentlabel\theequation \def\@tempb{#1}
\ifx\@tempb\empty\else\label{#1}\fi
\refstepcounter{alphaequation}
\let\@currentlabel\thealphaequation
\global\@eqnswtrue\global\@eqcnt=0 \tabskip\@centering\let\\=\@eqncr
$$\halign to \displaywidth\bgroup \@eqnsel\hskip\@centering
$\displaystyle\tabskip\z@{##}$&\global\@eqcnt\@ne
\hskip2\arraycolsep\hfil${##}$\hfil& \global\@eqcnt\tw@\hskip2\arraycolsep
$\displaystyle\tabskip\z@{##}$\hfil
\tabskip\@centering&\llap{##}\tabskip\z@\cr}
\def\endeqnsystem{\@@eqncr\egroup$$\global\@ignoretrue} \makeatother

\def\PAMELA{{\sc Pamela}}
\def\FERMI{{\sc Fermi}}
\def\HESS{{\sc Hess}}
\def\AMS{{\sc Ams-02}}

\begin{document}

\begin{flushleft}
\scriptsize{CERN-PH-TH/2012-171 \hfill 
SACLAY--T12/048 \hfill
LAPTH-038/13}
\end{flushleft}

\vspace{-0.02cm}

\begin{center}

{\Huge\bf Bremsstrahlung gamma rays\\[3mm] from light Dark Matter}

\medskip
\bigskip\color{black}\vspace{0.4cm}

{
{\large\bf Marco Cirelli}$^{\,a}$,
{\large\bf Pasquale D. Serpico}$^{\,b}$,
{\large\bf Gabrijela Zaharijas}$^{\,c}$
}
\\[7mm]	

{\it $^a$ \href{http://ipht.cea.fr/en/index.php}{Institut de Physique Th\'eorique}, CNRS, URA 2306 \& CEA/Saclay,\\ 
	F-91191 Gif-sur-Yvette, France}\\[3mm]
{\it $^b$ Laboratoire de Physique Th{\'e}orique d' Annecy-le-Vieux (\href{http://lapth.cnrs.fr/}{LAPTh}), Univ. de Savoie, CNRS, B.P.110, Annecy-le-Vieux F-74941, France}\\[3mm]
{\it $^c$ Abdus Salam International Centre for Theoretical Physics (\href{http://www.ictp.it}{ICTP}),\\
	Strada Costiera 11, 34151 Trieste, Italy}

\end{center}

\bigskip

\centerline{\large\bf Abstract}
\begin{quote}
\color{black}
\large
We discuss the often-neglected role of bremsstrahlung processes on the interstellar gas in computing indirect signatures of Dark Matter (DM) annihilation in the Galaxy, particularly for light DM candidates in the phenomenologically interesting ${\cal O}$(10) GeV mass range. Especially from directions close to  the Galactic Plane, the $\gamma$-ray spectrum is altered via two effects: directly, by the photons emitted in the bremsstrahlung process by energetic electrons which are among the DM annihilation byproducts; indirectly,  by the modification of the same electron spectrum, due to the additional energy loss process in the diffusion-loss equation (e.g. the resulting inverse Compton emission is altered). 
We quantify the importance of the bremsstrahlung emission in the GeV energy range, showing that it is sometimes the dominant component of the $\gamma$-ray spectrum. We also find that, in regions in which bremsstrahlung dominates energy losses, the related $\gamma$-ray emission is only moderately sensitive to possible large variations in the gas density. 
Still, we stress that, for computing precise spectra in the (sub-)GeV range, it is important to obtain a reliable description of the Galaxy gas distribution as well as to compute self-consistently  the $\gamma$-ray emission and the solution to the diffusion-loss equation.
For example, these are crucial issues to quantify and interpret meaningfully $\gamma$-ray map `residuals' in the inner Galaxy.
\end{quote}

\newpage

\section{Introduction}

Nowadays, Weakly Interacting Massive Particle Dark Matter (WIMP DM) candidates are routinely searched for using gamma ray emission from their pair-annihilation in the Galaxy; on the other hand, it has become clearer and clearer that this search has to be performed against poorly known astrophysical signals and non-negligible uncertainties in astrophysical backgrounds. It is thus of paramount importance to produce the sharpest possible prediction for a given DM candidate and to assess realistically the signal uncertainty. This is especially useful if independent hints for DM candidates should come from colliders and/or direct detection experiments.  It is also well known that astrophysics determines not only the foregrounds for indirect searches, but also enters  primarily in the signal normalization (as a function of the direction), a notable example being the unknown cuspiness of the dark matter halo towards the Galactic Center (GC).  

Less obvious and appreciated is the fact that astrophysical parameters 
may significantly shape {\it the energy spectrum itself}. This refers to the fact that DM generates both `prompt' and `secondary' photons:  prompt emission results from the fragmentation and decay of
different modes (gauge bosons, heavy fermions, higgses, \ldots) in addition to the loop-suppressed monochromatic photon channels. This part of the emission is controlled by particle physics, i.e. its spectrum is not affected by astrophysics, so that in a multi-messenger perspective one can infer some properties of the DM candidate by combining different signals.
On the other hand, secondary emission results from the energy losses of electrons and positrons emitted from the annihilations. 
This component {\it is} determined by astrophysics, essentially because the losses are determined by the environment.

As an example of important secondary radiation, one can cite the Inverse Compton Scattering (ICS) emission associated to heavy DM candidates (especially if characterized by `leptophilic' final states) which, in the recent years,  has been shown to be of crucial importance for constraints/diagnostic power~\cite{"old"ICS}. This class of models has been invoked particularly in scenarios attempting to explain the `excesses' in positron and electron data (as measured by \PAMELA, \FERMI, \HESS\ and, most recently, \AMS) as due to Dark Matter annihilation~\cite{Cirelli:2012tf}.
One may think that this aspect is somewhat `pathological' of heavy and leptophilic DM candidates: since they mostly produce leptons, and since these leptons are emitted at TeV energy scale, well above the GeV scale most easily accessible by an instrument like the LAT on board of \FERMI,  astrophysical details may somewhat matter in the `tail' of the  energy spectrum.

Here we show that this phenomenon is in fact more generic, by presenting another class of candidates for which the astrophysical effects in determining the gamma spectrum
are non-negligible. Namely, we shall concentrate on light WIMPs (${\cal O}$(10) GeV), which are motivated by several hints of putative signals in underground, direct detection experiments~\cite{DDhints}---most recently by the  $3\sigma$ level excess reported by {\sc Cdms}-II~\cite{Agnese:2013rvf}---as well as some gamma-ray and radio excesses~\cite{INDhints}, see also~\cite{Hooper:2012ft} for a review. We shall argue that, in presence of significant branching ratios into leptons, the role of the bremsstrahlung process in the e.m. field of the atoms of the interstellar gas alone can significantly alter the steady state electron/positron population  as well as the resulting overall gamma emission in the (sub-)GeV range~\footnote{{Notice  that, while the effect for light DM is expected to be most relevant (i.e., it affects secondaries carrying the bulk of the energy available), even for larger DM masses the effects that we study play a r\^ole for the tail of low energy electrons falling in the regime of energies (see below) where bremsstrahlung is important. Although for heavier candidates in the 0.1-1 TeV mass range prompt $\gamma$ radiation usually provides a better diagnostic tool, for specific applications to relatively low energies this caveat should be kept in mind.}}. Additionally, these effects are position-dependent (so that the DM spectrum is quite non-universal within the Galaxy) and are entangled with other astrophysical parameters (like the size of the diffusive halo). Further, we note that (similarly to interstellar radiation field and magnetic fields) the gas distribution especially in the inner regions of the Galaxy is only poorly known; unfortunately this uncertainty is often not taken into account self-consistently in the available numerical codes, which may cause some bias in predicting correlated signals in multi-wavelength and multi-messenger analyses. 

Given the clearly large number of astrophysical issues involved, we refrain from providing a complete assessment of the uncertainties on all the relevant astrophysical parameters. Our goals are rather: i) To stress that these poorly accounted effects introduce an uncertainty not only in the foreground estimates, but also in the energy spectrum associated to (recently popular) dark matter candidates. We will do so by referring to
recent astrophysics literature, although by no means in an exhaustive way. ii) To raise awareness on the fact that according to the actual numerical implementation of the energy losses and emissivities, further uncertainties may be induced by the numerical prescriptions used to predict the signals. This is particularly important if gas distributions models different from the `baseline' ones (i.e. used in the solution of diffusion/loss equation) are naively implemented in the emissivity calculations only.

{For example, any quantitative interpretation of `residual' gamma-ray as well as radio emission (in particular if in direction close to the Galactic Center or plane, as discussed in~\cite{INDhints} and~\cite{Hooper:2012ft}}) cannot prescind from accounting for the bremsstrahlung contribution:  For a given model of astrophysical foreground injection, uncertainties in the gas distribution as well as technical approximations in the solutions of the relevant equations might alter the estimate of these residuals  (see~\cite{YusefZadeh:2012nh} for a discussion of related aspects). 
 Even if significant `excesses' are confirmed, the bremsstrahlung contribution can change the interpretation of the results (for example altering the best fit annihilation cross section, mass, and preferred final states). This is not limited to dark matter models, of course, but applies also to other astrophysical models such as pulsars or black hole contributions (see e.g.~\cite{Abazajian:2010zy,Boyarsky:2010dr,Gordon:2013vta}).

{Note also that correcting $\gamma$-ray maps for DM purposes by merely removing excess emissions following some gas map template does not fully account for the neglected effect considered here. The spatial dependence of the electron injection, energy losses (and possibly diffusion properties) makes the steady state electron spectrum and thus the projected bremsstrahlung intensity map morphologically different with respect to a projected gas density map. Some modeling thus appears necessary to gauge the importance of the effects pointed out in this paper.}

This study emphasizes therefore once again the need to improve our knowledge of astrophysical parameters in order to sharpen our chances to see signals from DM and to take with a grain of salt the often claimed `universality' of the DM gamma spectrum as an important signature.

\bigskip

The rest of this paper is organized as follows: In Sec.~\ref{computation} we introduce the various  formulae and ingredients used in the computation; in particular,
after recalling the particle physics input we describe the gas maps used (subsec.~\ref{sec:gasmaps}) and the method to compute the electron spectrum (subsec.~\ref{sec:electronspectra}) 
Note that for computations we primarily use {\sc Pppc4dmid} \cite{PPPC4DMID}, modified to include  bremsstrahlung emission and energy losses. In addition, each step of the calculation is cross checked by using the fully numerical {\sc Galprop} code \cite{Galprop}. Results are presented in Sec.~\ref{sec:results}, while Sec.~\ref{sec:conclusions} is devoted to conclusions.

\section{Computation}\label{computation}

\begin{figure}[t]
\begin{center}
\includegraphics[width=0.48\textwidth]{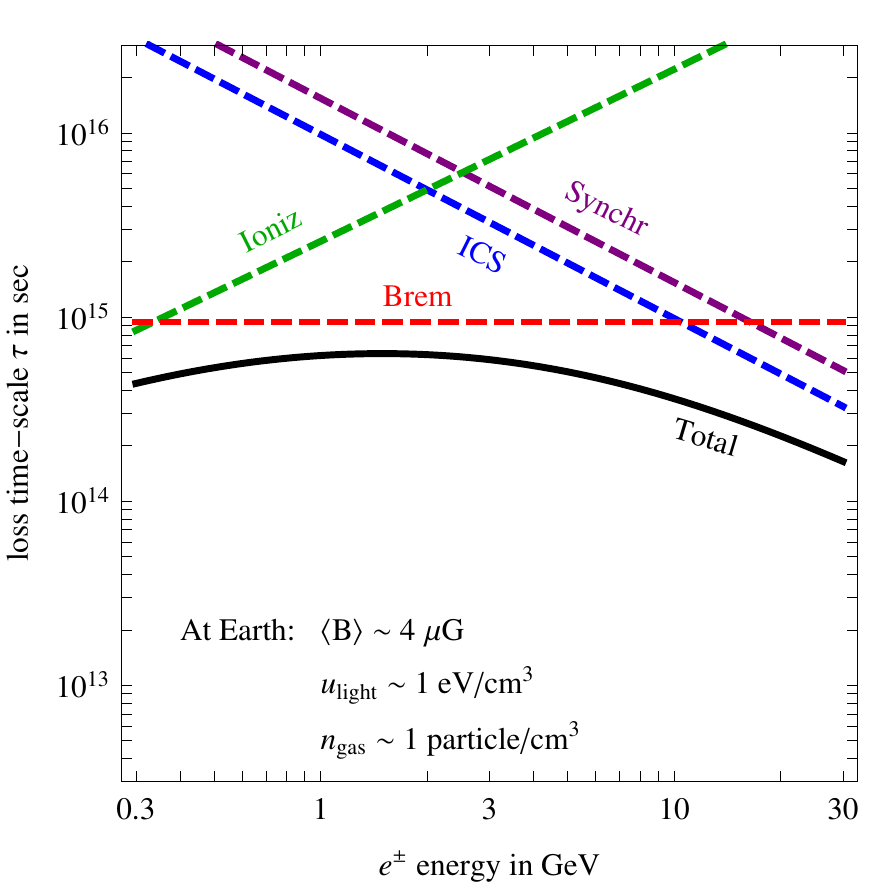}\quad
\includegraphics[width=0.48\textwidth]{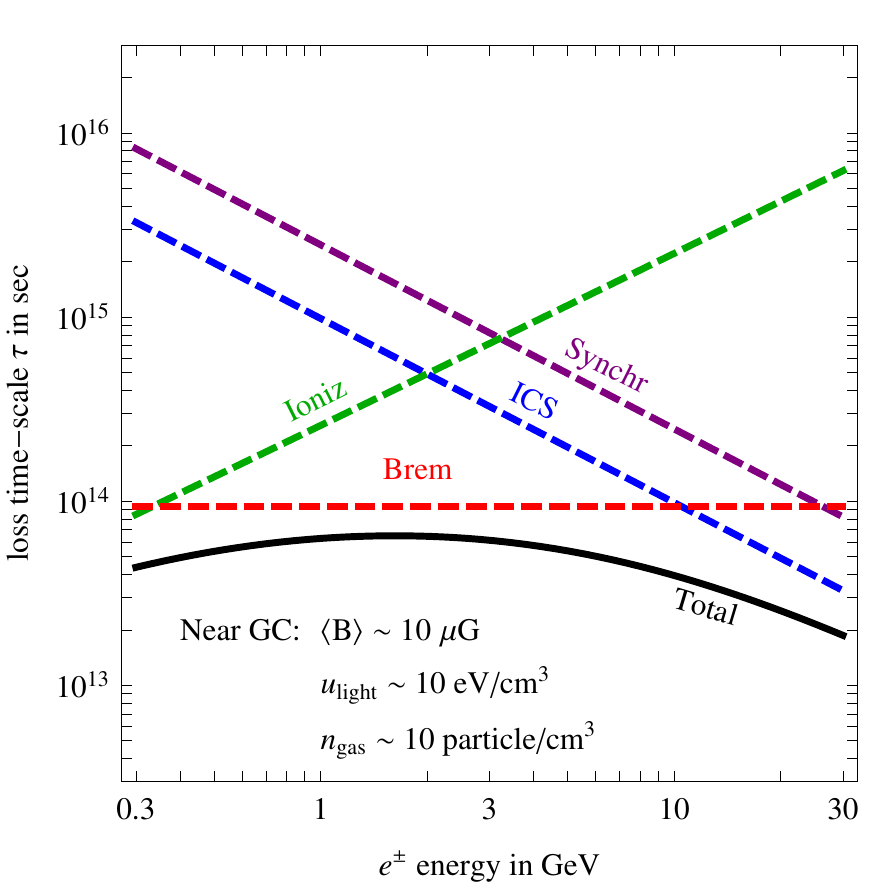}
\caption{\em \small \label{fig:enlosses} Typical {\bfseries  timescales of energy losses} of an electron (or positron) due to different processes, at the location of the Earth (left) and close to the galactic center (right). Here $\langle B \rangle$ is the average density of the magnetic field (relevant for synchrotron), $u_{\rm light}$ is an indicative density of interstellar radiation field (relevant for ICS) and $n_{\rm gas}$ is the density of the interstellar gas (relevant for bremsstrahlung; near the GC we choose 10 p/cm$^3$ for illustration: densities can go from $\mathcal{O}$(1) to $\mathcal{O}$(100) or more, see the discussion in Sec.~\ref{sec:gasmaps}). {We also plot an indicative timescale for diffusion: we choose `MED' parameters, hence the choice of a typical distance of 4 kpc (the vertical size of the diffusive halo in the `MED' configuration).}}
\end{center}
\end{figure}

Before delving into the formalism, let us briefly review the qualitative picture. Imagine a cell located at position $\vec x$ in the Galaxy (in cylindrical coordinates, since we always assume cylindrical symmetry, $ \vec x \to (r,z)$, with $r$ the projection on the galactic disk of the distance from the galactic center and $z$ the elevation with respect to the disk). The electrons and positrons present in the cell have been produced by DM annihilations (in the cell itself and) in a large surrounding volume and then diffused into the cell itself. During diffusion they have lost energy mainly by 1) ICS emission, 2) synchrotron radiation and 3) bremsstrahlung in the e.m. field of the interstellar gas atoms, which is our main interest.\footnote{Another process of energy loss is ionization on the interstellar gas, which is however relevant at very low $e^\pm$ energy, as shown in Fig.~\ref{fig:enlosses}: we will neglect it for the sake of simplicity from now on.} Fig.~\ref{fig:enlosses} shows the relative importance of these processes by comparing the characteristic timescales of energy loss, as a function of the $e^\pm$ energy. As apparent, for a range below $\sim$10 GeV, bremsstrahlung is the {\it dominant} process for typical environmental conditions and therefore should not be neglected.
 
In turn, the $\gamma$-ray emission of the cell consists of: 1) the prompt $\gamma$-rays from DM annihilations, 2) the ICS $\gamma$-rays from the energetic $e^\pm$ contained in the cell hitting the local background light and 3) the bremsstrahlung $\gamma$-rays from those same electrons hitting the gas in the cell (synchrotron emission falls in the radio/microwave range and will not be of our interest as a signal). Hence, there are essentially two ways in which the $\gamma$-ray signal is affected by bremsstrahlung: 1) via the impact that the additional energy loss has on $\Phi_{e^\pm}$, the steady state spectrum of $e^\pm$ later undergoing ICS; 2) via the additional emission process. In most DM $\gamma$-ray studies both aspects are neglected.

We now move to discuss bremsstrahlung by recalling some basic formul\ae, while details can be found in standard references such as~\cite{Blumenthal:1970gc,Schlickeiser}. We shall assume everywhere relativistic incident electrons and positrons (hence their speed is $c=1$), having energy $\gg 20$ MeV.

\bigskip

Schematically, one can write the gamma-ray emission ${\cal E}$ due to bremsstrahlung from a cell located at $\vec x = (r,z)$ as the integral over all possible $e^\pm$ energies $E_{e^\pm}$ giving origin to a photon of energy $E_\gamma$
\begin{equation}
\label{eq:flux}
\frac{{\rm d} {\cal E}_{\gamma, \rm brem}(\vec x)}{{\rm d} E_\gamma} = \sum_i n_i(\vec x) \int_{E_L} {\rm d}E_{e^\pm} \ 2 \frac{{\rm d} \Phi_{e^\pm}(\vec x)}{{\rm d} E_{e^\pm}} \cdot \frac{{\rm d} \sigma_i}{{\rm d}E_\gamma}
\end{equation}
where $n$ is the gas density (the index $i$ runs on all possible species) and the factor of 2 accounts for electrons {\it and} positrons ($\Phi_{e^\pm}$ represents the steady state electron {\it or} positron flux). $E_L={\rm max}(E_\gamma,E_{\rm min})$, $E_{\rm min}$ being the minimum energy cutoff.

The main particle physics input is the differential cross-section ${\rm d} \sigma_i/{\rm d} E_\gamma$ for bremsstrah-lung over the atomic, ionic or molecular species $i$.
This is usually written as
\begin{equation}
\label{eq:sigma}
\frac{{\rm d} \sigma_i(E_{e^\pm},E_\gamma)}{{\rm d} E_\gamma}=\frac{3\,\alpha_{\rm em}\sigma_T}{8\pi\,E_\gamma}\left\{\left[1+\left(1-\frac{E_\gamma}{E_{e^\pm}}\right)^2\right]\phi^i_1-\frac{2}{3}\left(1-\frac{E_\gamma}{E_{e^\pm}}\right)\phi^i_2\right\}\,,
\end{equation}
where $\sigma_{\rm T} = 8\pi \,\alpha_{\rm em}^2/(3\,m_e^2)$ is the Thomson cross section  and $\phi^i_{1,2}$ are scattering functions dependent on the properties of the scattering system. For example, for a completely ionized gas plasma consisting of a species of charge $Z$ one has
\begin{equation}
\label{eq:phiplasma}
\phi^{\rm ion}_1(E_{e^\pm},E_\gamma)=\phi^{\rm ion}_2(E_{e^\pm},E_\gamma)=4(Z^2+Z)\left\{\log\left[\frac{2E_{e^\pm}}{m\,c^2}\left(\frac{E_{e^\pm}-E_\gamma}{E_\gamma}\right)\right]-\frac{1}{2}\right\}\,,
\end{equation}
 where the $Z$ addendum in the pre-factor accounts for the electrons which neutralize the ion positive charges (one always assumes global charge neutrality). On the other hand, for atomic neutral matter the scattering functions have a more complicated dependence, which is usually parameterized in terms of the  quantity
$\Delta = \frac{E_\gamma m_e}{4 \alpha_{\rm em} E_{e^\pm} (E_{e^\pm}-E_\gamma)}$. For the ultra-relativistic regime one is usually interested in, one basically cares for the limit $\Delta \to 0$ for which these functions are constant and take the following numerical values: 
\begin{equation}
\label{eq:phiSS}
\begin{aligned}
\phi^{\rm H}_{1}(\Delta = 0) & \equiv \phi^{\rm H}_{1, {\rm ss}} = 45.79,\\ 
\phi^{\rm H}_{2}(\Delta = 0) & \equiv \phi^{\rm H}_{2, {\rm ss}} = 44.46, \\
\phi^{\rm He}_{1}(\Delta = 0) & \equiv \phi^{\rm He}_{1, {\rm ss}} = 134.60, \\
\phi^{\rm He}_{2}(\Delta = 0) & \equiv \phi^{\rm He}_{2, {\rm ss}} = 131.40, \\
\phi^{{\rm H}_2}_{(1,2)}(\Delta = 0) &\simeq 2\, \phi^{\rm H}_{(1,2), {\rm ss}}, 
\end{aligned}
\end{equation}
where we just listed the species of interest for parameterizing interstellar medium constituents (see Sec.~\ref{sec:gasmaps}). The subscript $_{\rm ss}$ in this notation refers to the fact that this regime is usually called `strong-shielding' because the atomic nucleus is screened by the bound electrons and the impinging $e^\pm$ have to force the shield.

The gamma-ray flux ${\rm d}\Phi_{\gamma, \rm brem}/{\rm d}E_{\gamma}$ from a given direction is then obtained by summing the emissions from all the cells located along the that direction, i.e. performing the integral along the line of sight of eq.~(\ref{eq:flux}).

\medskip

Similarly, one can derive the $e^\pm$ energy loss due to bremsstrahlung as the integral over all the emitted photon energies
\begin{equation}
\label{eq:enloss}
b_{\rm brem} \equiv -\frac{{\rm d}E_{e^\pm}}{{\rm d}t} = c\,\sum_i n_i\int_0^{E_{e^\pm}} {\rm d} E_\gamma\,E_\gamma\,\frac{{\rm d} \sigma_i}{{\rm d} E_\gamma}\,.
\end{equation}
For scattering on ionized matter (`weak shielding' regime), one has\,\footnote{This expression is actually valid for any energy $E_{e^\pm}$ of the incident $e^\pm$ and not only for the relativistic case, which is however the only one of interest.}
\begin{equation}
\label{eq:enlossWS}
b_{\rm brem}^{\rm ion} = \alpha_{\rm em} \frac{3\, \sigma_{\rm T}}{2\pi} n_i \ Z(Z+1) \left(\log 2 \ \frac{E_{e^\pm}}{m_e}-\frac{1}{3} \right) E_{e^\pm}\,.
\end{equation}
On the other hand, for relativistic scattering on atomic neutral matter (`strong-shielding') in the $\Delta\to 0$ limit, one can write
\begin{equation}
\label{eq:enlossSS}
b_{\rm brem}^{\rm neut} =\alpha_{\rm em} \frac{3\, \sigma_{\rm T}}{8\pi} n_i \left(\frac{4}{3} \phi^i_{1,{\rm ss}} -\frac{1}{3} \phi^i_{2,{\rm ss}} \right) E_{e^\pm}\,.
\end{equation}
Note that, at leading order, energy losses are linearly dependent from $E_{e^\pm}$. A further logarithmic dependence arises for scattering in ionized medium, while a small additional energy dependence is also found in neutral medium if one accounts for the effect of finite $\Delta$. In practice, neutral atomic gas constitutes the dominant component (as we will see in the next subsection) and the scattering is always in relativistic regime so that eq.s (\ref{eq:enlossSS}) and (\ref{eq:phiSS}) would be enough for our purposes. We will anyway use the full description in the numerical computations.

\subsection{Interstellar gas maps}
\label{sec:gasmaps}

Interstellar gas is composed mainly of hydrogen, which is dominantly found in atomic (HI) and molecular (H$_2$) forms. While cold H$_2$ clouds are tightly concentrated within the Galactic Plane, HI has a more significant scale height. Ionized hydrogen (HII) is subdominant in mass, but due to its large scale height, it is important to consider it when modeling large scale emission (which will not be the case in this work). Helium (He) has an abundance of 11\% with respect to H~\cite{Asplund:2004eu} and it is assumed that its distribution follows that of interstellar hydrogen. Heavier elements and dust make less than 1\% and we neglect them in the following.

HI maps are in general derived from the measurements of the 21-cm spectral line, emitted in transitions between the atomic hydrogen $S_2$ ground state levels split by the hyperfine structure.  
H$_2$, on the other hand cannot be directly observed in emission, as it is has no permitted lines at radio frequencies. It is therefore traced {\it indirectly} by using emission lines of CO (typically the 2.6 mm one caused by the $J=1\rightarrow 0$ transition of the CO excited by the collisions with the H$_2$ molecules). There is a considerable uncertainty in relating  the column densities of measured CO and more abundant H$_2$ and it is captured by their proportionality factor $X_{\rm CO}$ (see for example~\cite{FermiLAT:2012aa} or~\cite{Cholis:2011un}). This is one of the main sources of uncertainty in obtaining H$_2$ maps, while the error on the so-called spin temperature is the source of dominant uncertainty for HI maps.

In what follows we will implement in the {\sc Pppc4dmid} code 2D analytical gas model maps as used in {\sc Galprop} and described in \cite{Moskalenko:2001ya}. This choice facilitates our comparison of the results from the two codes we use. 

In detail, for the radial distribution of the HI we use Gordon \& Burton~\cite{GordonBurton} (Table 1), while the vertical distribution is from Dickey \& Lockman (1990) \cite{Dickey&Lockman} and Cox {\it et al.} (1986) \cite{Cox}. The models used for H$_2$ maps are the ones of  Bronfman {\it et al.}~\cite{Bronfman} for $1.5$ kpc $\leq r \leq$ 10 kpc, and that from Wouterloot {\it et al.}~\cite{Wouterloot} for $r \geq$ 10 kpc, augmented with the Ferri\`ere {\it et al.}~\cite{Ferriere} model for $r \leq$ 1.5 kpc, see \cite{Moskalenko:2001ya}.  
{For $X_{\rm CO}$ we adopt the constant value $1.9 \cdot 10^{20}$ mol cm$^{-2}$/(K km s$^{-1}$), although other approaches are possible, see fig. 25 of~\cite{FermiLAT:2012aa} or eq. (10) of~\cite{Cholis:2011un}.}
For the HII maps we use \cite{Gaensler}. Finally, atomic and ionized He maps are rescaled from the ones of HI and HII respectively, with the 0.11 factor cited above.
Fig.~\ref{fig:gasmaps} illustrates the resulting maps along selected slices of the Galaxy, relevant to our computations.
 
There are competing and, in some cases, more recent determination of gas densities, as for example the hydrogen maps of~\cite{Nakanishi:2003eb},~\cite{Pohl:2007dz} and~\cite{Tavakoli:2012jx}, but the differences that this choice induces are not critical for our purposes.
Indeed, in this work we care primarily about the {\it relative} differences between models in which we vary the amount of bremsstrahlung energy losses and therefore the exact choice of the maps is of secondary importance.

\begin{figure}[t]
\label{fig:gasmaps}
\begin{center}
\includegraphics[width=0.31\textwidth]{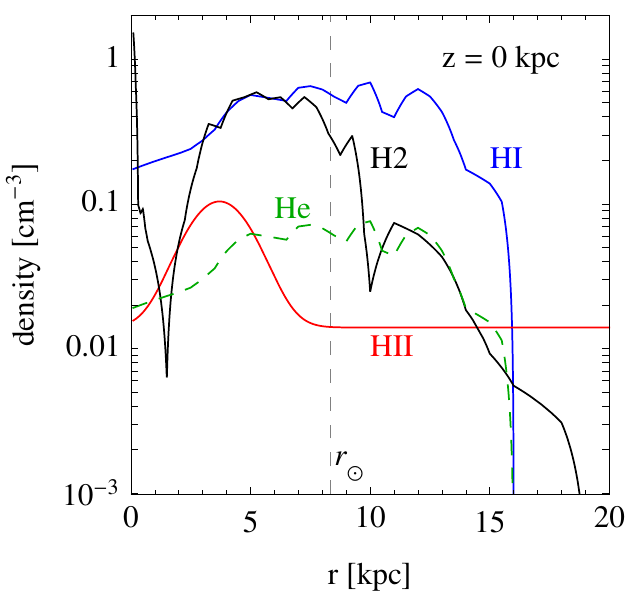}\quad
\includegraphics[width=0.31\textwidth]{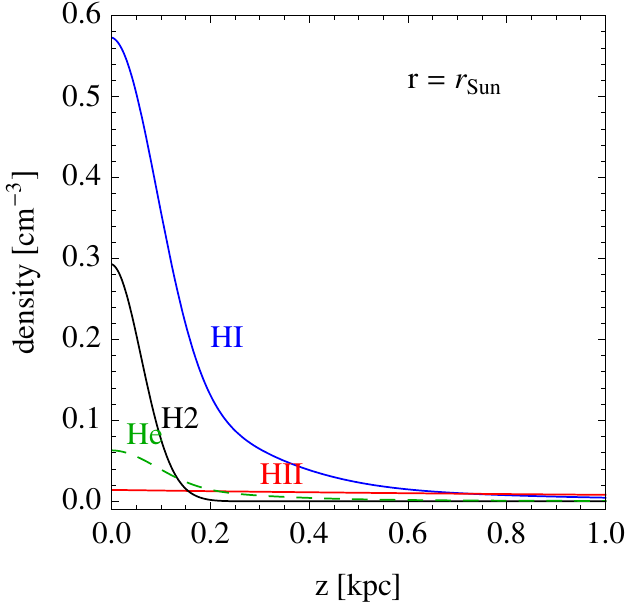}\quad
\includegraphics[width=0.31\textwidth]{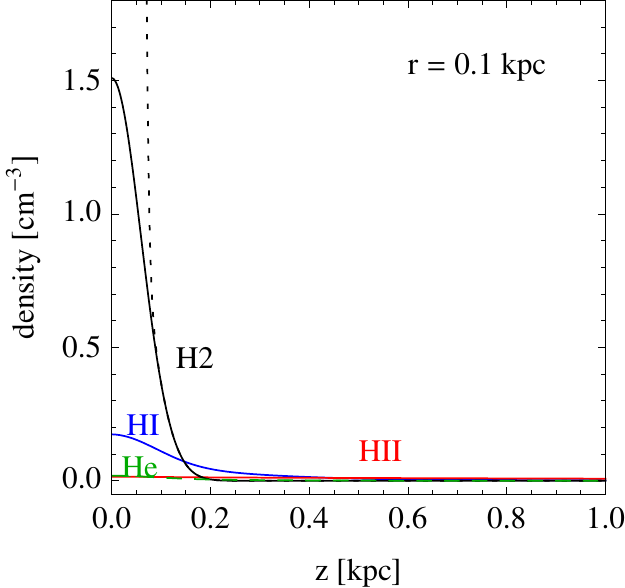}
\caption{\small \em \label{fig:gasmaps} Slices of the large scale {\bfseries gas density maps} that we adopt; left: as a function of $r$ in the galactic disk; center: as a function of $z$ above and below the Sun; right: as a function of $z$ above and below a point close to the GC. The dotted line traces the enhanced {\rm H2} density employed in the `realistic $n_{\rm gas}$ model of the GC, see text.}
\end{center}
\end{figure}

\bigskip

A side comment is now in order. In general, the coarse 2D maps ($\Delta r \sim \Delta z\geq 0.5$ kpc) listed above are not accurate enough to interpret the $\gamma$-ray sky maps with the angular resolution of the \FERMI-LAT instrument ($\Delta l, ~\Delta b~\approx ~0.1^\circ$, where $l, ~b$ are Galactic longitude and latitude, respectively). The approach implemented for example in the {\sc Galprop} code is then to calculate the $\gamma$-ray fluxes along the line of sight $ds$ by taking advantage of the high resolution $l,~b$ maps based on radio-astronomical surveys: the 21 cm survey~\cite{Kaberla}, which has a resolution of $0.5^\circ$ and the 115 GHz Center for Astrophysics survey of CO~\cite{Dame}, with a resolution of $0.125^\circ$ to $0.25^\circ$. The integral of the emission is therefore corrected for the observed column densities of the high resolution maps ($N_{{\rm HI},k}$ and $W_{{\rm CO},k}$ below) while maintaining the model-based variation within the centro-galactic rings $k$ ($n_{\rm HI}$ and $n_{{\rm H}_2}$), as:
\begin{equation}\label{eq:hiresmaps}
\frac{{\rm d}\Phi_{\gamma, \rm brem}}{{\rm d}E_{\gamma}}=\Sigma _i \frac{N_{{\rm HI},i}+2~X_{{\rm CO},i}W_{{\rm CO},i}}{\int{n_{\rm HI}+2n_{{\rm H}_2}~ds}}\times \int{\frac{{\rm d} {\cal E}_{\gamma, \rm brem}}{{\rm d} E_\gamma} ~(n_{\rm HI}+2n_{{\rm H}_2})~ds}
\end{equation}
The individual conversion factors $X_{{\rm CO},k}$, analogous to the one discussed above, are free parameters in each ring and are then obtained a posteriori in a fit to the gamma-ray data. For more information on this procedure see \cite{FermiLAT:2012aa}.
The potential problems with this approach is that it mixes together uncertainties in the background density of the gas and uncertainties in the inhomogeneity of the high-energy population of CR, which are both energy dependent. For example, protons and electrons are not distributed equally due to energy losses (even if they were emitted equally). If maps are calibrated as above, in a region where $p$-$p$ dominates, that could introduce a systematic effect in the region where electron induced processes are relevant.
In addition it has been noted~\cite{Isabelle} that not all gas is accounted for using the above tracers and that dust emission provides a good tracer of the gas (so called `dark gas'). A dust correction is applied to the gas maps in {\sc Galprop} to account for this effect leading to a much better agreement with the observed gamma ray emission. While this involved procedure may be useful to optimize a fiducial background map for some specific application, here we choose to use, self-consistently, the same gas maps for the computation of the energy losses and for the emission, in order to grasp directly the physics involved.

\bigskip

Another point to discuss, still related to the accuracy of the maps but much more relevant for our purposes, concerns the region of the inner Galaxy. While the maps that we use are in general adequate on the large scales of the Galaxy, 
their validity is questionable e.g. in regions close to the galactic center (typically at $\leq~2^\circ\sim200$ pc scales). This area hosts active star forming regions and several structures of dense gas, in particular: 
\begin{itemize}
\item[$\circ$] The Central Molecular Zone (CMZ): made up dominantly of molecular gas, placed in an asymmetric layer extending (in projection) out to $r\sim 200$ pc (or, more precisely, $r\leq250$ pc at positive longitudes and $r\leq150$ pc at negative longitudes).
Ferri\`ere {\it et al.}~\cite{Ferriere} estimate that densities in the CMZ reach $\langle {\rm n_{H2}}\rangle_{\rm CMZ,max}\approx 150 \ {\rm cm}^{-3}$ (but keep in mind uncertainties like the one of a factor $\sim 2$ induced by $X_{\rm CO}$~\cite{Ferriere}). 

\item[$\circ$] The Circum-Nuclear Ring (CNR): region of the inner 1 to 3 pc believed to be part of an accretion disk around the central massive black hole. Its gas content was studied, under the approximation of a smooth and axisymmetric ring, in \cite{Ferriere:2012ni}, finding gas densities of $\langle {\rm n_{H2}} \rangle_{\rm CNR} = 2.2\times10^5$ cm$^{-3}$.

\end{itemize}

The geometry of these regions is quite involved, and they do have a certain  level of clumpiness (see for example~\cite{Launhardt}). Yet, it is clear that within the CMZ volume,  average densities are of the order $\sim 100$ cm$^{-3}$, with possibly higher densities being reached at smaller scales~\cite{Ferriere:2012ni}. The resolution in the galacto-centric distance of the high resolution gas maps described above (eq.~(\ref{eq:hiresmaps})) is limited to $\geq$ 1 kpc by the finite non-circular motions of the gas traced by these surveys as well as internal velocity dispersions of molecular clouds, therefore variations in the gas densities at this small scales are largely unresolved~\footnote{Note that the innermost annulus (0-1.5 kpc) is entirely enclosed within the interpolated region. For that reason the HI gas maps used in {\sc Galprop} are made by assuming the innermost annulus contains 60\% more gas than its neighboring annulus, \cite{FermiLAT:2012aa}.}. For comparison, the maps illustrated in Fig.~\ref{fig:gasmaps} (which are also used in {\sc Pppc4dmid}) reach a maximal value of the H$_2$ density at the GC of only 3 cm$^{-3}$. 

Based on the discussion above, in the following we will consider two different models for the region around the GC:
\begin{itemize}
\item[-] The `coarse-grained' maps of Fig.~\ref{fig:gasmaps}, with their unrealistically low gas density of ${\cal O}(1) \ {\rm cm}^{-3}$, which we will label as `{\bf coarse-gd $n_{\rm gas}$}'.
\item[-] The addition of a crude modeling for the above-mentioned structures (in particular the CMZ) by adding a cylindric slab around the GC of radial size of  200 pc and scale height of 50 pc within which the gas density is enlarged 100 times with respect to the coarse-grained maps, reaching values of ${\cal O}(100) \ {\rm cm}^{-3}$. We will label this case as `{\bf realistic $n_{\rm gas}$}'. Needless to say, this is only a toy model, but likely closer to the reality than the first one.
\end{itemize}
 It should be also noted that further uncertainties are probably induced by likely modification to the diffusion properties in this inner region, while for simplicity (and lack of empirical constraints) we stick to a homogeneous diffusion coefficient with properties identical to local ones.

\subsection{Electron spectra}
\label{sec:electronspectra}

Having discussed the astrophysical environment, the first step of the computation consists in determining the spectral density of $e^\pm$, injected by Dark Matter and subsequently propagated. 

{
The $e^\pm$ number density per unit energy, $f(t,\vec x,E)= dN_{e^\pm}/dE$,
obeys the diffusion-loss equation~\cite{SalatiCargese}:
\beq 
\label{eq:diffeq}
\frac{\partial f}{\partial t}-
\nabla\left( \mathcal{K}(E,\vec x) \nabla f \right) - \frac{\partial}{\partial E}\left( b(E,\vec x) f \right) = Q(E,\vec x)
\eeq
with diffusion coefficient function $\mathcal{K}(E,\vec x)$
and energy loss coefficient function $b(E,\vec x)$. 
They respectively describe transport through the turbulent magnetic fields and energy loss due to several processes. Notice that other terms would be present in a fully general diffusion-loss equation for Cosmic Rays, such as diffusive re-acceleration terms (describing the diffusion of CR particles in momentum space, due to their interactions on scattering centers that move in the Galaxy with an (Alfv\'en) velocity $V_a$) and convective terms. These are however negligible for $e^\pm$, see e.g.~\cite{SalatiCargese,Delahaye:2008ua}.
Eq.~(\ref{eq:diffeq}) is solved in a diffusive region with the shape of a solid flat cylinder that sandwiches the galactic plane, with height $2L$ in the $z$ direction and radius $R=20\,{\rm kpc}$ in the $r$ direction~\cite{DiffusionCylinder}. The location of the solar system corresponds to $\vec x  = (r_{\odot}, z_{\odot}) = (8.33\, {\rm kpc}, 0)$~\cite{rSun}.
Boundary conditions are imposed such that the $e^\pm$ density $f$ vanishes on the surface of the cylinder, outside of which electrons and positrons freely propagate and escape.

The solution of the diffusion-loss equation is determined using} the tools provided by the {\sc Pppc4dmid}~\cite{PPPC4DMID}, to which we refer for a {more complete} discussion. We just recall that $d\Phi_{e^\pm}/dE$, the differential flux of $e^\pm$ in each given point $\vec x$ of our Galaxy, can be written as
\beq
\label{eq:positronsflux}
\frac{{\rm d}\Phi_{e^\pm}}{{\rm d}E}(E,\vec{x}) =
\frac{c}{4\pi \, b(E,\vec x)}
\frac12 \left(\frac{\rho(\vec x)}{m_{\rm DM}}\right)^2 \sum_f \langle \sigma v \rangle_f \int_E^{M_{\rm DM}} {\rm d}E_{\rm s} \, \frac{{\rm d}N^f_{e^\pm}}{{\rm d}E}(E_{\rm s}) \, {I}(E,E_{\rm s},\vec{x})
\eeq
where $\rho(\vec x)$ is the DM halo profile, $m_{\rm DM}$ is the DM mass, $\langle \sigma v\rangle_f$ is the velocity averaged annihilation cross section into the channel with final state $f$ and ${\rm d}N^f_{e^\pm}/{\rm d}E$ is the $e^\pm$ injection spectrum for the same channel. $b(E,\vec x)$ is the energy loss coefficient function and $I(E,E_{\rm s},\vec x)$ are the generalized halo functions introduced in~\cite{PPPC4DMID}.
We improve on this treatment, however, in a number of ways: 1) we include of course in $b(E,\vec x)$ the energy losses for bremsstrahlung, as computed in eq.~(\ref{eq:enloss}); 2) we increase the accuracy of the halo functions in the region of the Galactic Center ($r \lesssim 0.5$ kpc) in order to be able to follow in detail the emission there; 3) we implement a refined map of the Interstellar Radiation Field (ISRF)~\footnote{{In particular we use the ISRF map from the {\sc Galprop} package: {\tiny 'MilkyWay DR0.5 DZ0.1 DPHI10 RMAX20 ZMAX5 galprop format.fits.'} For more details on how it was produced see \cite{FermiLAT:2012aa} and references therein.}} to compute ICS losses (and later the ICS emission).~\footnote{These improvements will be included in an upcoming release of the {\sc Pppc4dmid}~\cite{PPPC4DMsyn}.}
The results that we obtain are cross-checked with {\sc Galprop} under the same conditions, finding a reasonably good agreement: typically, the difference goes from the percent level to several tens of percent for extreme conditions; in the latter case, however, these are within the larger uncertainties that we wil discuss later. In general, what we aim at is a fully self-consistent calculation which correctly reproduces the main features of the results and, especially, the relative differences rather than the absolute values.

For the underlying density of DM we choose for definiteness an NFW profile with parameters as given in Figure 1 of~\cite{PPPC4DMID}. The $e^\pm$ diffusion/loss parameters are fixed, again for definiteness, at `MED', as given in Table 2 of~\cite{PPPC4DMID}. Changing these choices (see e.g.~\cite{Nesti:2013uwa} for more recent determinations of the DM halo parameters) would lead to modifications in the computed spectra but would not change the qualitative picture.

\section{Results}
\label{sec:results}

\subsection{Electron spectra}
\label{sec:electronspectraresults}

\begin{figure}[t]
\begin{center}
\includegraphics[width=0.31\textwidth]{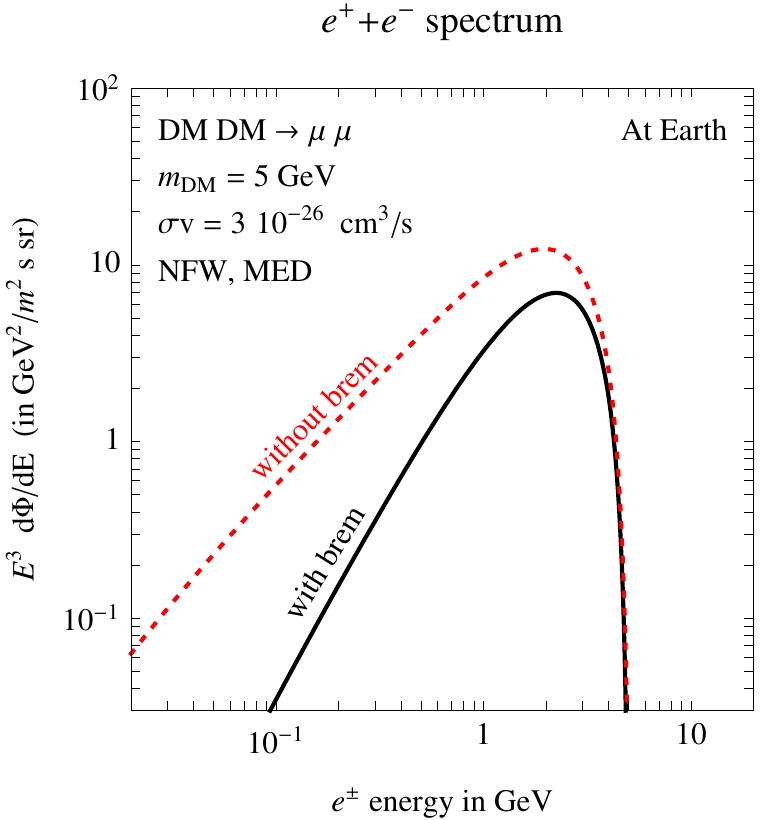}\quad
\includegraphics[width=0.31\textwidth]{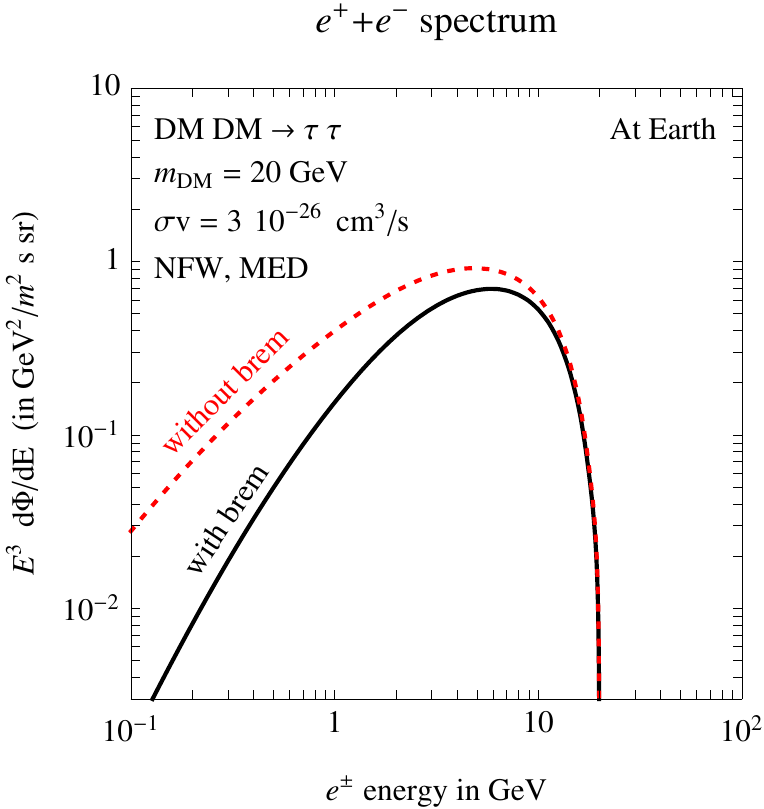}\quad
\includegraphics[width=0.31\textwidth]{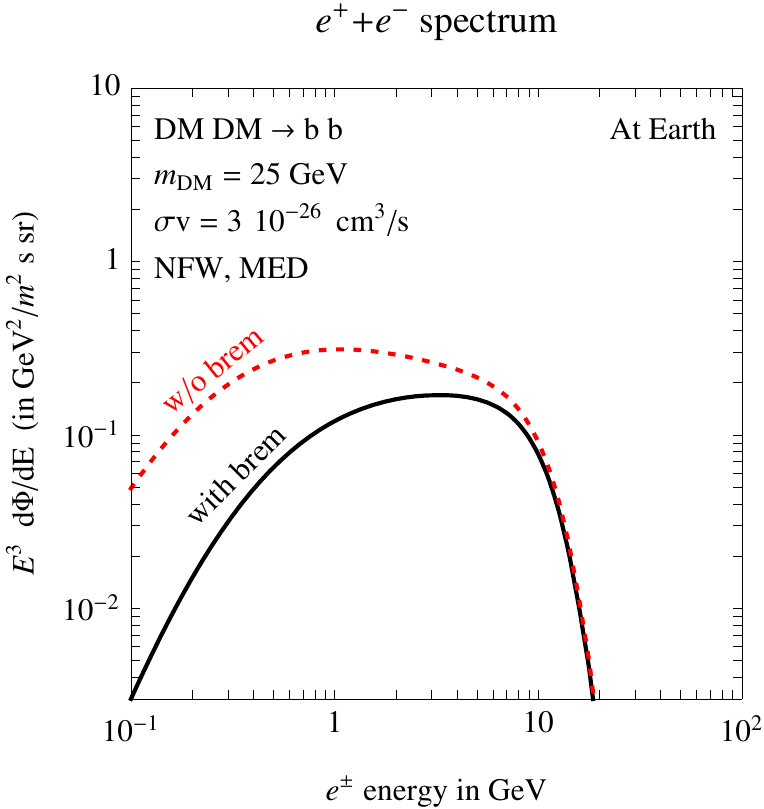}\\[0.5cm]
\includegraphics[width=0.31\textwidth]{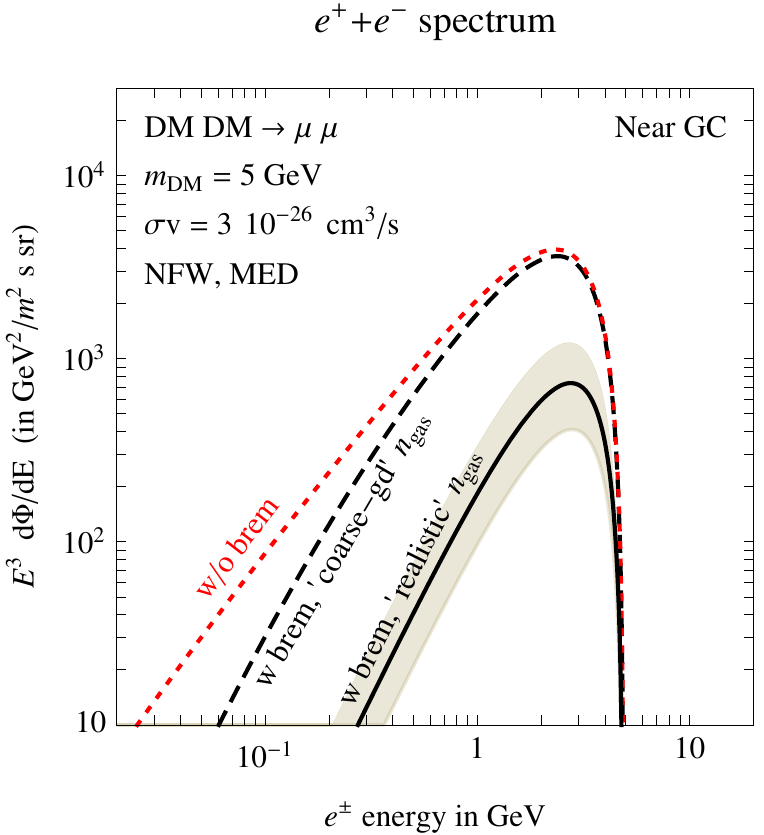}\quad
\includegraphics[width=0.31\textwidth]{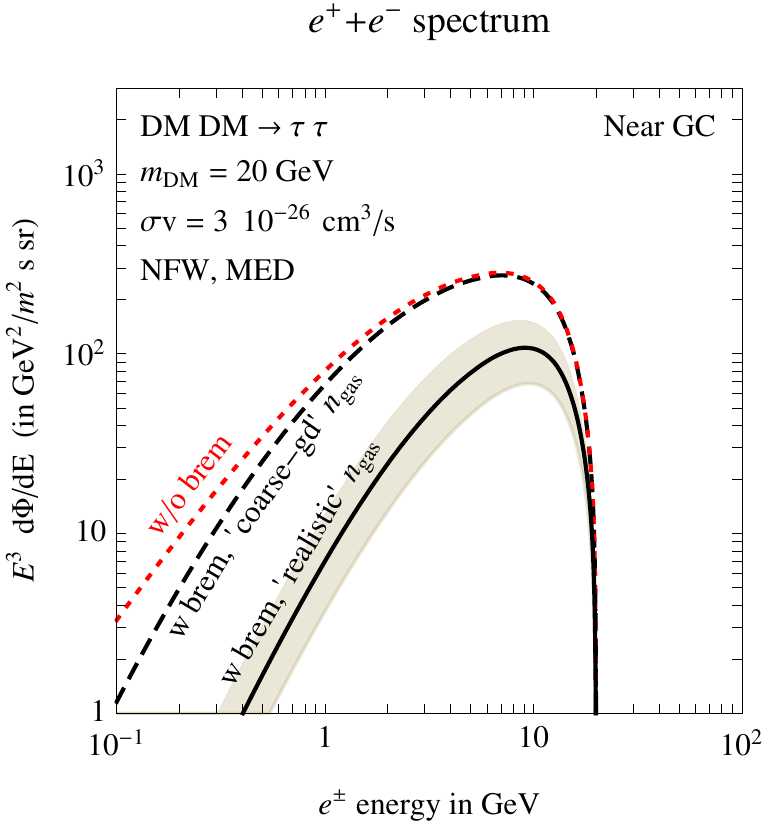}\quad
\includegraphics[width=0.31\textwidth]{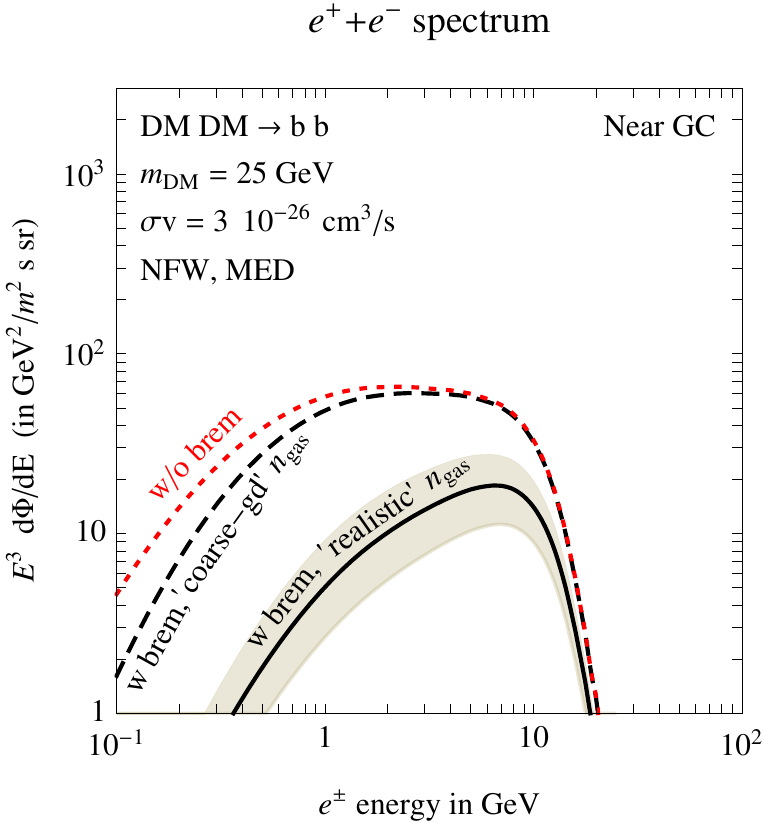}
\caption{ \em \small \label{fig:espectra} {\bfseries Electron + positron spectral density} at two locations in the Galaxy (at the Earth, top row, and near the GC at $(r,z) = (0.1,0) \, {\rm kpc}$, bottom row), for three specific DM models (columns). The dashed red line represents the spectrum one would have neglecting bremsstrahlung losses; the solid black line is the actual spectrum including bremsstrahlung; the dotted black line corresponds to enlarging the gas density to values realistic for the GC region (see Sec.~\ref{sec:gasmaps}).}
\end{center}
\end{figure}

In fig.~\ref{fig:espectra} we show the electron spectra at the Earth's position $(r,z = 8.33, 0)$ kpc and at a location $(r,z = 0.1, 0)$ kpc that we consider as representative of the conditions close to the Galactic Center and which is inside the high gas density region discussed above. Indeed, for this location we compute the spectra in three cases: neglecting bremsstrahlung, including bremsstrahlung with the coarse grained gas maps of fig.~\ref{fig:gasmaps} (`coarse-gd $n_{\rm gas}$') and including bremsstrahlung with a gas density enlarged by 100 times, typical of the CMZ (`realistic $n_{\rm gas}$'). In addition, the band around the `realistic $n_{\rm gas}$' case corresponds to an uncertainty of a factor 2 in $n_{\rm gas}$, representative of the uncertainties in the CMZ like those related to $X_{\rm CO}$ discussed above.

We choose three representative DM models: annihilation into $\mu^+\mu^-$ of a 5 GeV particle, annihilation into $\tau^+\tau^-$ of a 20 GeV particle and annihilation into $b \bar b$ of a 25 GeV particle. With these choices, the bulk of the injected $e^\pm$ falls in the range of energies (few GeV) where bremsstrahlung is important. 
We always assume a thermal annihilation cross section of $3 \ 10^{-26} {\rm cm}^3/{\rm s}$. 

These plots show immediately that:
\begin{itemize}
\item[{\small $\blacktriangleright$}] Neglecting bremsstrahlung losses altogether leads to a sizable error at the peak and in the low-energy tail of the spectrum. 
\item[{\small $\blacktriangleright$}] A proper modeling of the gas density distribution near the GC is probably crucial. The difference between the `coarse grained' maps and those including a modelling for the CMZ 
can easily change the resulting spectra by almost one order of magnitude.
\item[{\small $\blacktriangleright$}] In turn, the uncertainty on the actual gas density of the GC region has a sizable impact on the normalization of the spectrum (shaded band).
\item[{\small $\blacktriangleright$}] One can expect that if one were to adopt a simple rescaling of the emissivity in the GC region (with electron spectra still computed with coarse grained maps --the procedure discussed in sec.~\ref{sec:gasmaps}, which we do not adopt--) one would be led to significant errors in the estimated gamma-ray flux, since it is the steady state electron population itself to be affected.
\end{itemize}

\begin{figure}[t]
\begin{center}
\includegraphics[width=0.31\textwidth]{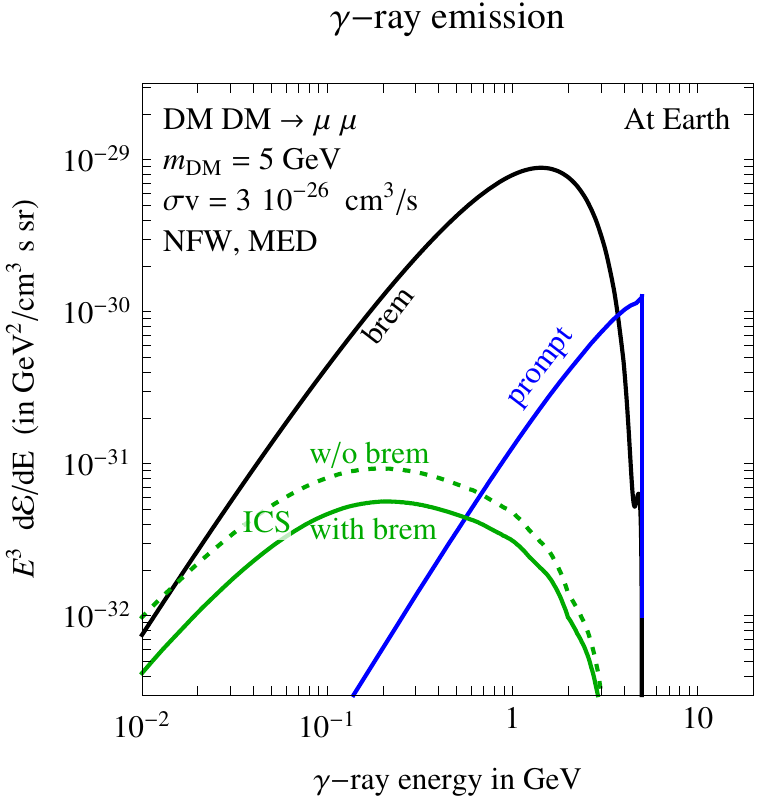} \quad 
\includegraphics[width=0.31\textwidth]{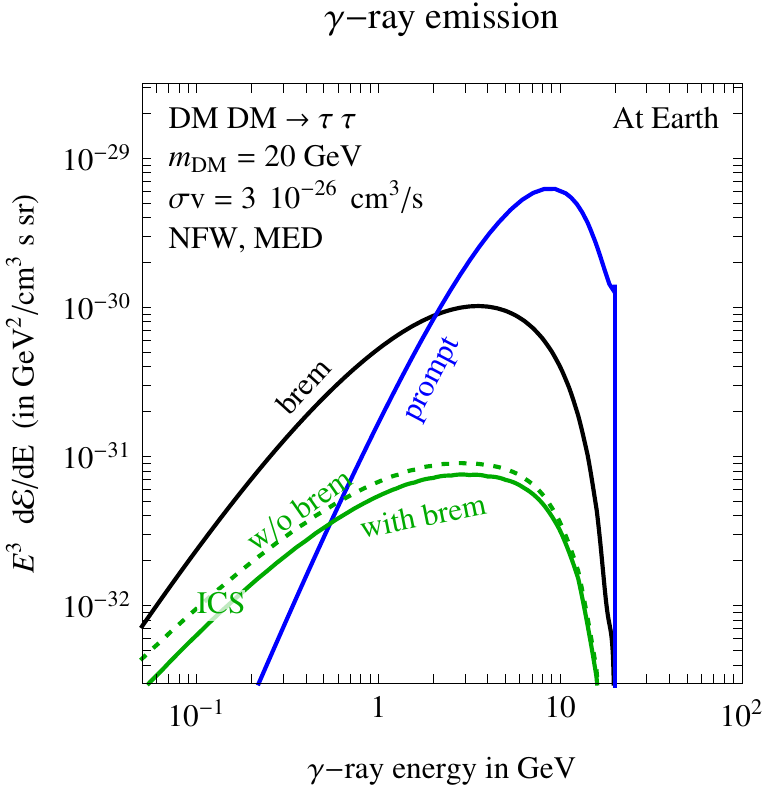} \quad 
\includegraphics[width=0.31\textwidth]{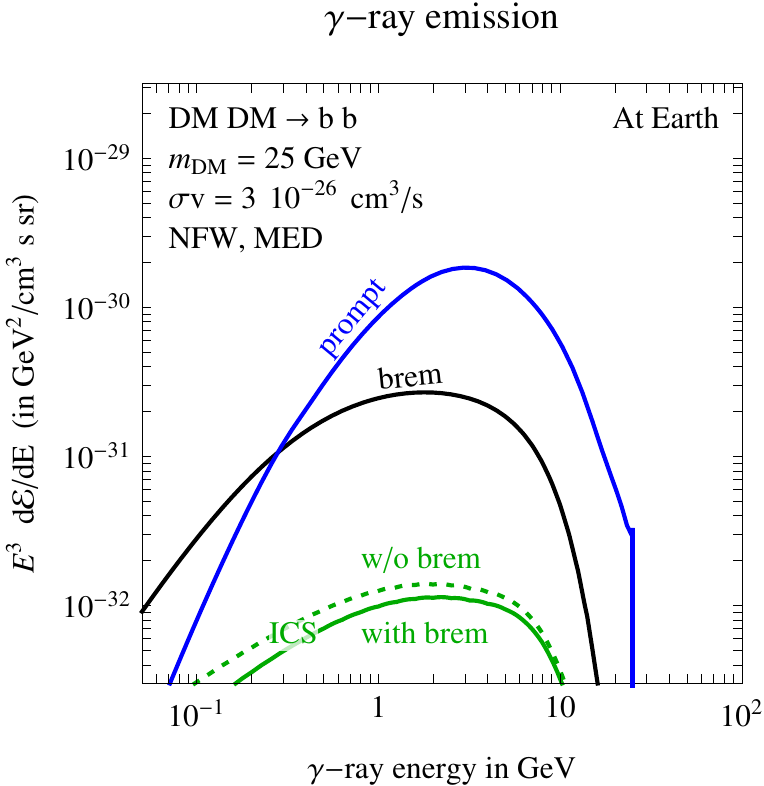} \\[0.3cm]
\includegraphics[width=0.31\textwidth]{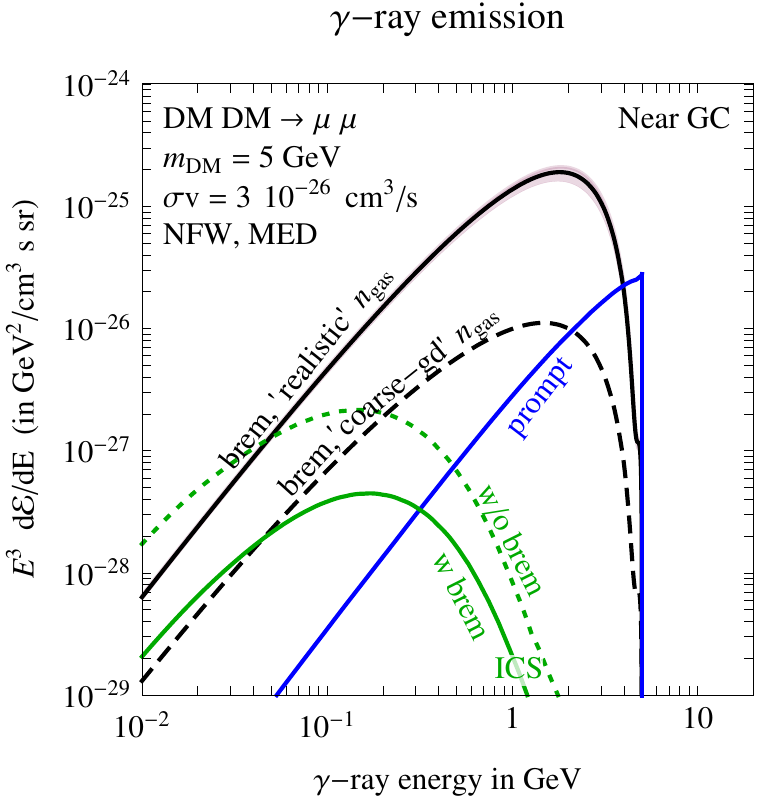} \quad 
\includegraphics[width=0.31\textwidth]{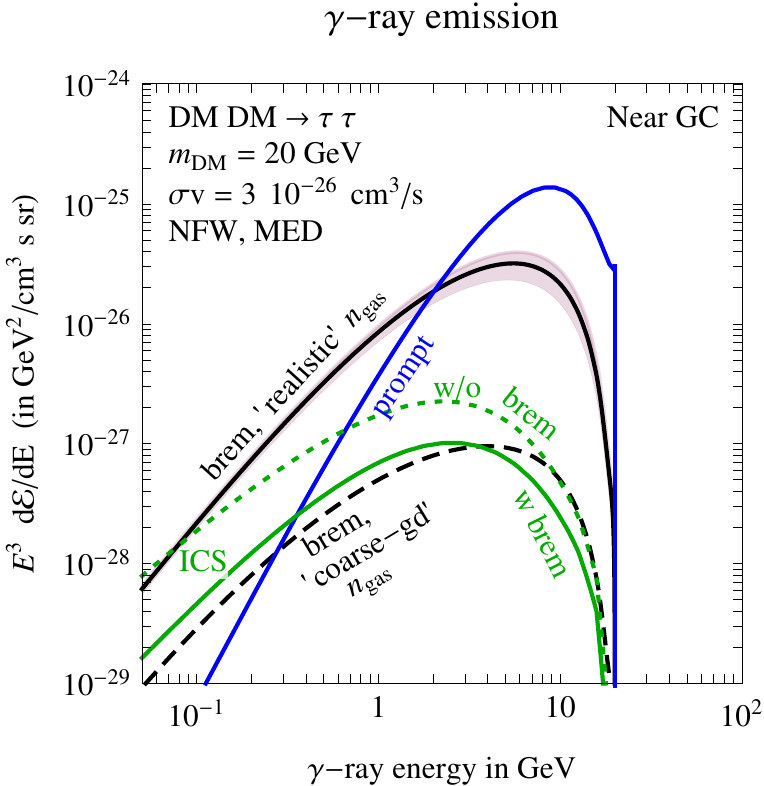} \quad 
\includegraphics[width=0.31\textwidth]{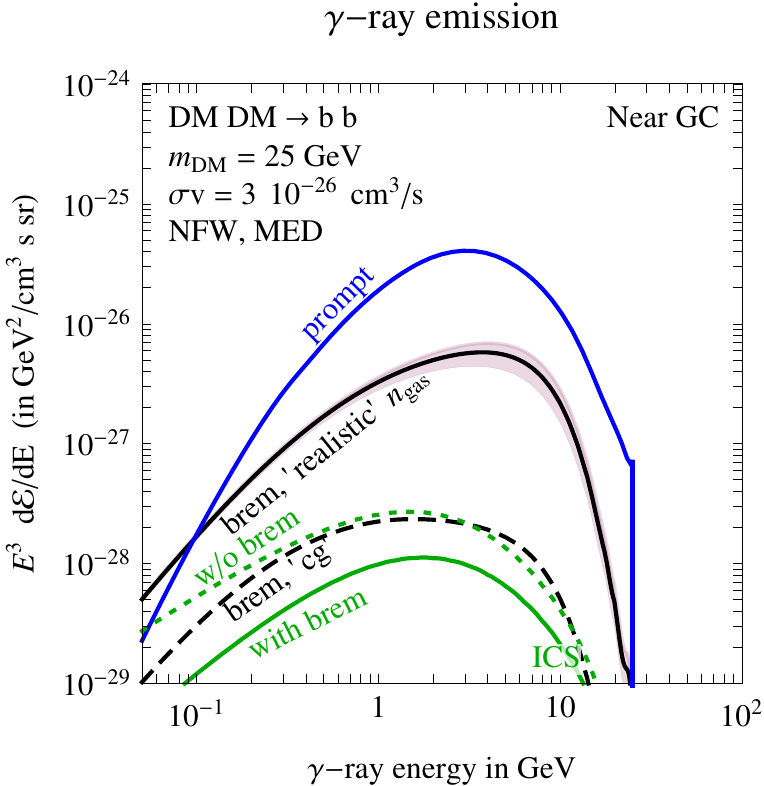}
\caption{\em \small \label{fig:spectra} The different components of the {\bfseries gamma-ray spectrum from DM}, at the same two locations as in fig.~\ref{fig:espectra} and for the same three specific DM models we consider (columns). We show the prompt contribution, the bremsstrahlung contribution (also assuming large gas densities at the GC, see text) and the ICS contribution (computed both including and neglecting bremsstrahlung energy losses).}
\end{center}
\end{figure}

\subsection{Gamma-ray spectra}

We can now proceed to compute the $\gamma$-ray spectra associated to the same DM models introduced above. In fig.~\ref{fig:spectra} we show the differential emissions ${\rm d}{\cal E}/{\rm d}E$ of the unit volume cells
({\it number of photons per unit time, energy, solid angle and volume, rescaled by $E^3$}) located at the Earth and close to the GC. In each case we show the different components: prompt (taken directly from~\cite{PPPC4DMID}), bremsstrahlung (computed according to eq.~(\ref{eq:flux})) and ICS (computed as discussed in~\cite{PPPC4DMID}). 
Several comments are in order. 
\begin{itemize}
\item[{\small $\blacktriangleright$}]
First and foremost, it is evident that bremsstrahlung gives a very important or---depending on the energy---even dominant contribution to the total $\gamma$-ray flux. It is therefore crucial to include it in order to have a reliable prediction of the spectrum. This is particularly true for the leptonic channels $\mu^+\mu^-$ and $\tau^+\tau^-$, and to a lesser extent for the hadronic $b \bar b$ channel.\\ 

\item[{\small $\blacktriangleright$}] Second, we see that the ICS component is also affected by the inclusion of bremsstrahlung energy losses, which modify the underlying electron spectra as discussed in the previous Subsection. 

\item[{\small $\blacktriangleright$}] Third, we see that the relative proportion of the different components is position dependent, as the bremsstrahlung (and the ICS) emissions depend on the environmental gas (and background light) density and composition, making the DM gamma-ray spectrum far from universal within the Galaxy.

\item[{\small $\blacktriangleright$}] Finally, the large variation of the bremsstrahlung component with and without the adoption of realistic gas maps at the GC illustrates the important impact of the local astrophysical uncertainties on the determination of the gamma-ray spectra. On the other hand, the precise value of the gas density in the CMZ is less important than for the $e^\pm$ spectra (i.e. the uncertainty band corresponding to varying $n_{\rm gas}$ by  a factor 2 is here relatively thinner). This is because a large $n_{\rm gas}$ suppresses on one side the $e^\pm$ steady state spectrum via bremsstrahlung energy losses but also, on the other side, boosts bremsstrahlung $\gamma$-ray emission, so that the two effects somewhat compensate, as long as bremsstrahlung is the dominant energy loss process. 
In particular we find that changes in the gas density in the range from 25 cm$^{-3}$ to 400 cm$^{-3}$ result in a $-$35\% to +15\% uncertainty band with respect to our  benchmark value for the $\mu^+\mu^-$ channel. The band illustrates rather robustly the estimate of the uncertainty, but we stress that a detailed knowledge of the inner Galaxy gas density would be required for a finer determination.
Needless to say, a firm determination may still prove crucial for morphological studies, to establish if one is truly in a regime dominated by the bremsstrahlung losses, or to gauge the impact on other emissions such as the inverse Compton ones. For example, a significant clumpiness or variation in the scale-height in the distribution of the gas may break the above mentioned balance of effects and lead to larger (more spectacular) variations in the signal.
\end{itemize}

\begin{figure}[t]
\begin{center}
\includegraphics[width=0.306\textwidth]{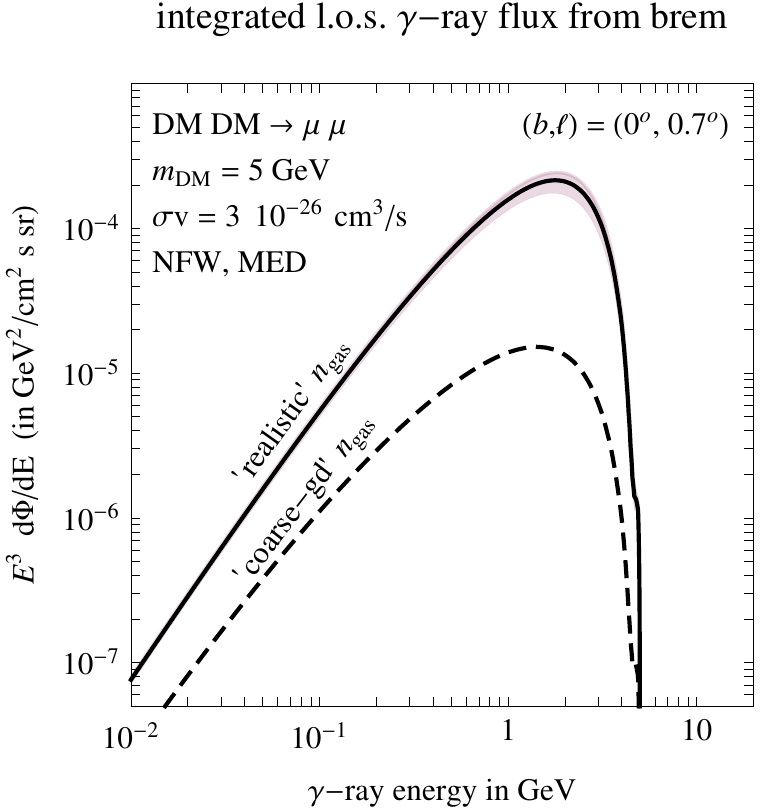}\quad
\includegraphics[width=0.31\textwidth]{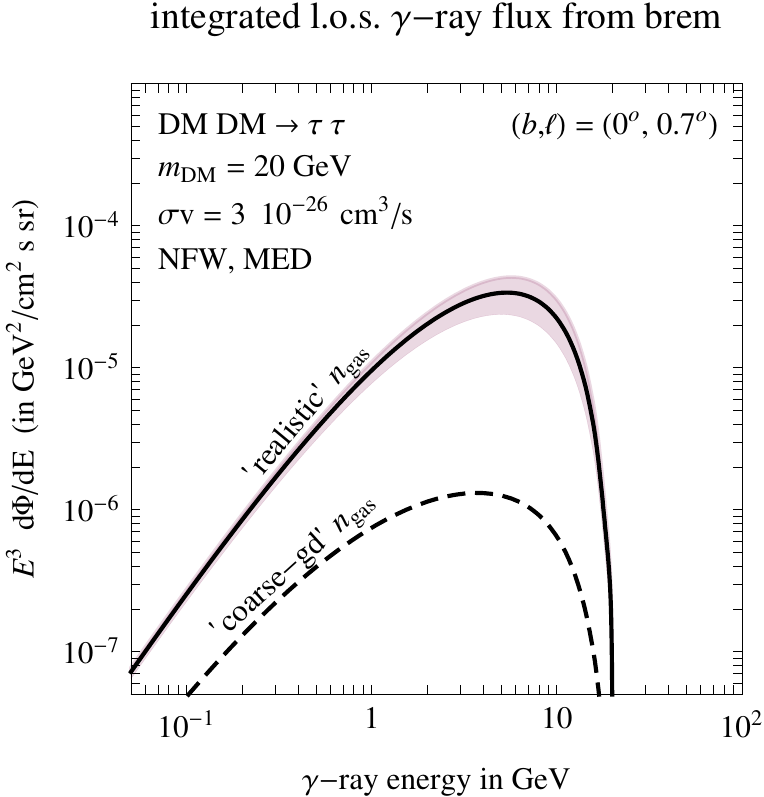}\quad
\includegraphics[width=0.31\textwidth]{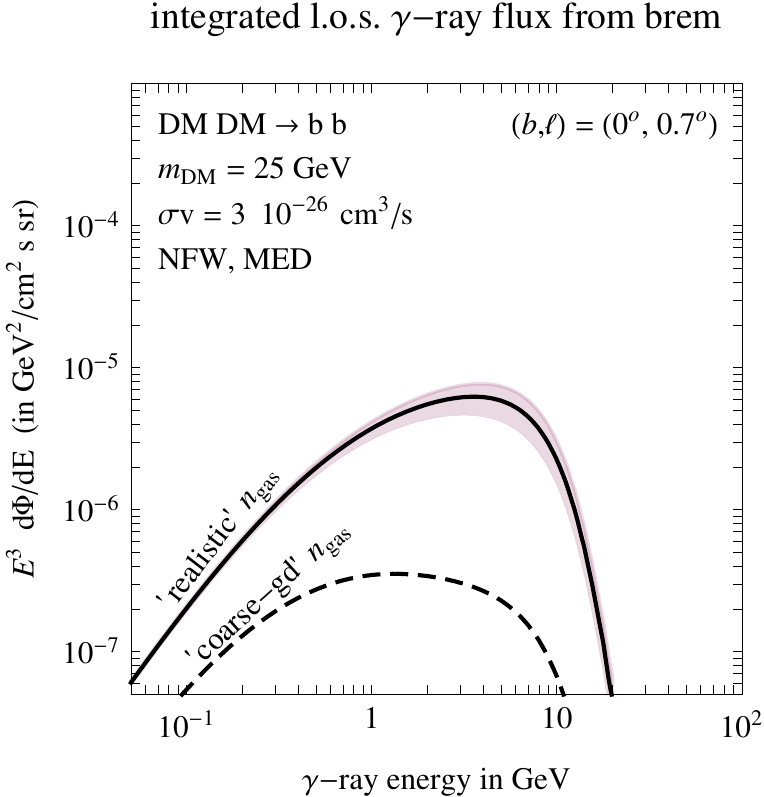}
\caption{ \em \small \label{fig:losspectra} The {\bfseries gamma ray spectrum} from bremsstrahlung, integrated along the line of sight, for the same three DM models that we consider. This refers to observation towards a point at latitude $b=0^\circ$ and longitude $\ell = 0.7^\circ$, i.e. a line of sight which passes $\simeq 0.1$ kpc ``to the East'' (or ``to the West'') of the GC.}
\end{center}
\end{figure}

In fig. \ref{fig:losspectra} we also show some examples of bremsstrahlung gamma-ray spectra integrated along the line of sight. These are quantities directly observable with gamma-ray instruments such as {\sc Fermi}-LAT.
We choose a l.o.s. lying inside the galactic plane ($b=0^\circ$) and at a longitude $\ell = 0.7^\circ$ with respect to the direction towards the Galactic Centre. The difference between the spectra computed with and without the realistic $n_{\rm gas}$ illustrates that including a fine description of the gas distribution in the GC region is crucial for the flux predictions, since the dominant emission comes indeed from the inner region. Although we did not perform an extensive analysis of errors in astrophysical parameters, it is clear that  the uncertainties in the central location of the Galaxy largely impact the prediction even for the integrated signal (here the bands have the same meaning as before, i.e. they correspond to the `minimal' uncertainty of order a factor of 2 such as the one following from errors in the CO to H conversion). {In Table \ref{tab:relerror} we explicitly show the relative error induced by neglecting the bremsstrahlung emission\footnote{Strictly speaking, we omit Inverse Compton contribution at the denominator, when gauging the total flux.  For the cases considered, as shown in Fig.~4, the IC contribution is either sub-leading with respect to bremsstrahlung or prompt contributions at the higher energies, or at most comparable to the bremsstrahlung contribution at lowest energies. Since however it suffer from other uncertainties and its inclusion would not change qualitatively the results, for the sake of clarity we prefer not to include it in the table.} at three energies of interest (0.1, 1 and 3 GeV), and show for comparison the relative error bar of the Fermi LAT data (as taken from \cite{FermiLAT:2012aa}) at those energies. } {As the Table shows, the impact of bremsstrahlung is larger (or even far larger) than the uncertainty in {\sc Fermi}-LAT and cannot be dismissed.}

\begin{table}[t]
\begin{center}
\begin{tabular}{|c||c|c|c||c| }
\hline
 Energy [GeV] & $\mu^+\mu^-$, 5 GeV & $\tau^+\tau^-$, 20 GeV & $b~{\bar b}$, 25 GeV & {\sc Fermi}-LAT error \\
\hline \hline
 0.1 & 0.99 & 0.98& 0.66 & 0.1 \\
 1 & 0.99 & 0.82& 0.25 & 0.1 \\
 3 & 0.93 & 0.54& 0.20 & 0.15 \\
 \hline
\end{tabular}
\caption{\em \small Ratio of bremsstrahlung to the total $\gamma$-ray flux, at three energies of interest (0.1, 1 and 3 GeV) and for the three DM candidates we consider in the text. The last column shows the {\sc Fermi}-LAT relative error-bars, as taken from Figure 15 in~\cite{FermiLAT:2012aa} (note that the region considered in that work is $|b|\leq ~8^\circ$, $l\leq~80^\circ$, which is expected to have comparable but {\em slightly higher} overall error bars than the region closer to the Galactic Center).}
\label{tab:relerror}
\end{center}
\end{table}

\section{Conclusions}
\label{sec:conclusions}

In this work we have discussed the often-neglected role of bremsstrahlung processes\footnote{One notable exception is ref. \cite{Tavakoli:2013zva}, which appeared several weeks after the completion of this paper.} on the interstellar gas for computing indirect signatures of Dark Matter annihilation in the Galaxy, especially for light DM candidates in the phenomenologically interesting ${\cal O}$(10) GeV mass range. We found that both {\it direct} and {\it indirect} effects are important in altering the expectations of the gamma spectrum observable at the Earth (especially for directions along the Galactic Plane, which we focused on). By {\it direct} we mean here the photons emitted in the bremsstrahlung process by energetic electrons which are among the DM annihilation byproducts. By {\it indirect} effects we refer to 
all the others which are affected by the modification of the resulting electron spectrum, due to  bremsstrahlung energy losses in the diffusion-loss equation. Here we illustrated the case of the Inverse Compton emission,
but we mentioned that the radio signal is similarly affected. 

In general, the effects are quite large in the GeV and sub-GeV range, and no reliable prediction of the gamma spectrum can be obtained if bremsstrahlung is neglected. This is particularly important for leptonic final states, but not negligible also for final states containing quarks. Also, the effect is maximal for directions close to the Galactic plane, where gas density is maximal. 

For directions close to the Galactic Center, a further challenge is due to the poor knowledge of gas distribution (not to speak of diffusion properties). We showed that adopting a `coarse grained' model of the gas which ignores inner Galaxy structure such as the CMZ, or implementing a toy model for this gas distribution in the inner 200 pc characterized by a large gas density dramatically affects the results, by one order of magnitude or so.
Although the latter model is closer to being realistic, we pointed out that the true morphological distribution of the gas, its degree of clumpiness, the value of the $X_{\rm CO}$ conversion parameter in the inner Galaxy, etc. are all factors contributing to the uncertainty of the signal. 
We also explained that the effect of this uncertainty cannot simply be modeled by altering the `emissivity'  maps to compute the gamma emission given a certain electron spectrum. In fact, the equilibrium electron spectrum
itself (given a certain source, here DM) can be significantly altered depending on the actual gas distribution. 
{Conceptually, the 2D morphology of the gamma-ray sky at the GeV energies cannot be determined solely by the 2D projected gas density maps, no matter how high their resolution:  modeling correctly the 3D distribution is necessary if the gas presents large variations in relatively small scale (at the 100 pc level or smaller).} Additionally,
in order to gauge the impact of different gas distributions, a self-consistent treatment where
they are used both in the diffusion-loss equation and to compute the emissivity is required. We further demonstrate that in the regime in which energy losses are dominated by the bremsstrahlung processes (as it could be the case in the Galactic Center region) due to the competing effects that high gas density has on the electron spectra and the gamma ray emission, large (eight fold) variations in the gas density have only a moderate overall impact on the bremsstrahlung gamma ray emission. 
{This would not be true, however, of other secondary emissions such as radio or IC ones,
which do not scale proportionally to the gas density.}

{Also it is worth clarifying that, although the gas densities and related uncertainties in the gas maps are highest in the inner 200 pc of our Galaxy, the impact of bremsstrahlung energy losses depends on the nature of a problem and on the related data analysis procedure and should be evaluated appropriately  for any region of interest, particularly when considering energies $\leq$ 3 GeV. Typically, whenever gas distributions with large variations in the gas densities at spatial scales well below the kpc are present, the impact of bremsstrahlung emission (and its uncertainty) might be relevant and should be checked.}

Our work reinforces the need for improved astrophysical modeling in order to reliably predict DM signals. Also, it suggests that especially towards the Galactic center, any interpretation of gamma-ray `residuals' in the GeV range with respect to expectations based e.g. on {\sc Galprop} predictions is tricky, for two reasons: i) Attributing an excess to a given DM model is not trivial, since the predicted gamma ray signal cannot be predicted very reliably. ii) The exact nature of `residuals' may be quantitatively ill-defined, because they could be at least partially mimicked by leaving the baseline injection sources unchanged, but altering the propagation parameters
and emissivity maps. Estimating the uncertainty in residuals by changing emissivity maps or diffusion parameters within their range {\it independently} may prove incorrect.

In this sense, the insights gained here go beyond the sole application to the DM problem, and provide useful caveats in the complicated analysis of the Galactic Center region, which is however
extremely rich and interesting both for astrophysics and for searches of physics beyond the Standard Model.


\bigskip

{\small 
\paragraph{Acknowledgements}

We thank Ilias Cholis, Maryam Tavakoli and Martin Pohl. MC is particularly grateful to Ga\"elle Giesen and Marco Taoso for useful discussions.

\noindent This work is supported by the European Research Council ({\sc Erc}) under the EU Seventh Framework Programme (FP7/2007-2013) / {\sc Erc} Starting Grant (agreement n. 278234 - `{\sc NewDark}' project). It is also supported in part by the French national research agency {\sc Anr} under contracts {\sc Anr} 2010 {\sc Blanc} 041301 and  {\sc DMAstroLHC} (ANR-12-BS05-0006- 01), and by the EU ITN network {\sc Unilhc}. 
MC (PS) acknowledges the hospitality of the Institut d'Astrophysique de Paris (the Institut de Physique Th\'eorique de Saclay), where part of this work was done. We also thank the Theory Unit of CERN for hospitality at different stages.
At LAPTh, this activity was developed coherently with the research axes supported by the Labex grant {\sc Enigmass}.}

\footnotesize
  

\end{document}